\documentclass[manuscript,twocolumn]{aastex701}

\usepackage{enumitem}

   \usepackage{booktabs}
   \usepackage{svg}

\shorttitle{\texttt{Freo Doctor}}
\shortauthors{Devillepoix et al.}
\setwatermarkfontsize{1in}
\graphicspath{{./}{figs/}}

\newtheorem{req}{Aspirational Requirement}

\begin{document}
\newcommand{\contrib}[1]{\textcolor{teal}{#1}}

\newcommand{\numb}[1]{\textcolor{orange}{#1}}
\newcommand{\source}[1]{\textsuperscript{\textcolor{blue}{[citation needed]}}\xspace}
\newcommand{\missnumber}{\numb{[NUMBER]}\xspace}
\newcommand{\checknumber}{\numb{[check number]}\xspace}

\title{\texttt{Freo Doctor}: Atmospheric Modelling for Meteorite Falls and Spacecraft Re-Entries}

\author[orcid=0000-0001-9226-1870,sname='Devillepoix']{Hadrien A. R. Devillepoix}
\affiliation{Space Science and Technology Centre, Curtin University, Boorloo/Perth WA, Australia}
\affiliation{International Centre for Radio Astronomy Research, Curtin University, Boorloo/Perth WA, Australia}
\email{hadrien.devillepoix@curtin.edu.au}

\author[orcid=0000-0003-2193-0867,sname='Cup\'ak']{Martin Cup\'ak}
\affiliation{Space Science and Technology Centre, Curtin University, Boorloo/Perth WA, Australia}
\affiliation{International Centre for Radio Astronomy Research, Curtin University, Boorloo/Perth WA, Australia}
\email{martin.cupak@curtin.edu.au}

\begin{abstract}
How much does the wind affect the path of meteorite falls?
We finely model the lower $\sim$30 km of the atmosphere using Weather Research and Forecasting open source tools at 1 km spatial resolution.
Models initialised at different times give different results, which can be used as a proxy for uncertainty.
We find that in most cases the differences on the ground positions are significant: median shift for a 1\,kg meteorite is 143\,m, doubling to 307\,m for a 10\,g rock, though these vary by over an order of magnitude between events.
The differences wind model choice makes on the ground are significantly larger than the typical uncertainty on meteoroid state vector obtained from bright flight observations of the fireball ($<100$\,m), and should be taken into account when predicting meteorite free-fall path to the ground.
Unsurprisingly the cases where we see the largest differences coincide with documented extreme weather events.
We also find that high spatial resolution models (1 vs. 3 km) tend to perform better.
We have successfully used these models to guide field teams to the location of 12 fallen meteorites after fireball observations.
We release as open data 1107 models we have calculated for 302 meteorite fall events and spacecraft re-entries around the world.
\end{abstract}

\section{Introduction} \label{sec:intro}

Fireball observations, either intentional using dedicated instrumentation or serendipitously captured by other sensors, allow to both determine orbital origins of the space rock sample and determine whether any meteorite survived entry.
To interpret the hypersonic bright phase of the atmospheric entry, a general atmospheric model such as \texttt{NRLMSISE-00} \citep{2002JGRA..107.1468P} is generally of sufficient fidelity.
Once the meteoroid decelerates to sub-hypersonic speed, a general model is no longer sufficient, as winds significantly affect the path of falling meteorites.
For a decimeter to meter scale meteoroid, the altitude at which this transition happens is typically between 10 and 35 km (most cases are between 20 and 30 km).
The precision of the numerical integration of a meteorite in dark flight depends on three things:
\begin{enumerate}
    \item \textbf{last known state vector}. With current observation techniques and bright flight modelling, the precision on the last known state vector is generally known to $<100$\,m.
    Some issue arise when the individual fragments undergo a sudden vector change right at the end of the bright flight \citep{2003M&PS...38.1023B,2024M&PS...59..927M}, however these problems ought to disappear in the future by employing better observation hardware.
    
    \item \textbf{physical properties of the meteorite}: mass, density, shape. These are difficult to estimate just from optical bright flight data and the subject of active research \citep{2020AJ....160...42B}. Fortunately, even when faced with poor knowledge of these physical parameters, the uncertainty in the size of the area to be searched for meteorites tends to grow linearly (instead of quadratically, as one might intuitively expect)\citep{2022PSJ.....3...44T}.
    
    \item \textbf{atmospheric conditions} along the fall path: wind speed and its direction, pressure, and temperature. This is the focus of the present work:
\begin{req}
\label{req:model}
Get wind direction and speed throughout the ground-30 km column at a given time of interest, and within 20\,km of a given location, 
accurate to a point where we can can predict the fall position of a 10\,g meteorite to within 50\,m.
\end{req}
\end{enumerate}

\subsection{What sources of atmospheric data are used for dark flight modelling?}

\subsubsection{Profiles as recorded by nearby radiosonde balloon flights}
Radiosonde balloons released for routine meteorological observations can go up to $\sim$ 35 km altitude before they burst.
They are a practical and straightforward way to get accurate observations of the direction and strength of the wind throughout the column of interest.
If a meteorological balloon is flown near in time and location of a meteorite fall, this is undoubtedly the best data source to use.
This is rarely the case for meteorite falls.
Meteorological offices have limited number of stations,
and typically only launch one balloon every 24 or 12 hours.
The positive effects of increasing the density (both in space and time) of radiosonde flights is well documented \citep{2014MWRv..142.1823P}.
However the trend over the last 30 years has been for national sounding programmes to reduce these activities.
In Australia, this corresponded with the switch from staffed to automated weather stations.
For instance, at Forrest airport, in the core of the present-day Desert Fireball Network \citep{Howie2017Howbuildcontinental}, the last Bureau of Meteorology balloon launch happened in 1995; the launches were then moved to nearby Eucla, but they also stopped there in 2012.
In 2015, Russia's Roshydromet agency cut its flights from two ascents per day to one at each station due to budgetary constraints \citep{ingleby2016radiosonde}.
And a similar thing is happening as of 2025 in the United States.
As of late, weather forecasting has recently been picked up by commercial entities, that have found the need to run their own commercial global soundings \citep{2026arXiv260202714S}.
This will likely improve weather forecasts overall, however it is unclear if these new raw observation data will be available in the public domain for research.
The first explicitly reported use of radiosonde as direct input for dark flight was by \citet{1978JRASC..72...15H} for the Innisfree meteorite fall, for which they used the nearest balloon flight from 3 hours before in Edmonton (140 km away).

\subsubsection{Models provided by national meteorological institutes}
Some meteor research groups collaborate with the national meteorological institutes in the countries where they operate.
This gives them access to data products that are not normally openly accessible, either because these are a paid service or they are impractically large to publish systematically.
For instance, the European Fireball Network get access to data products from the closed-source Aire Limitée Adaptation Dynamique Développement International (ALADIN) models run by the Czech Hydrometeorological Institute \citep[e.g.][]{2020M&PS...55..376S,Shrbeny:2026}.

\subsubsection{Global models}
Other works have reported using Global Forecast System (GFS) and European Centre for Medium-Range Weather Forecasts (ECMWF) operational analysis data as wind inputs \citep{2021MNRAS.503.3337M}.

\subsubsection{WRF modelling (this work)}
The Weather Research and Forecasting (WRF) model is an open-source mesoscale numerical weather prediction \citep{SkamarockEtAl2019}.
We have been calculating models after a meteorite fall 
observed by the Global Fireball Observatory \citep{2020P&SS..19105036D}.
Dark flight calculations using these models have led to the recovery of 12 meteorites so far (see discussion in Sec. \ref{sec:val_met}).
These models have also been used to do a posteriori dark flight analyses for some meteorite that were recovered before detailed fall area predictions became available, or using other models: \textit{Golden} \citep{Brown:2023}
and \textit{Saint-Pierre-le-Viger} \citep{2025NatAs...9.1624E}.
Finally, we have used the WRF models to study man-made re-entries to predict the fall location the debris from a \textit{Soyuz 2.1b} rocket re-entry over Australia \citep{Hatty:2026}.

In this paper, we describe what input data we use for calculating WRF models, the pre-processing steps, and the configuration used for the models integrations.
We focus the discussion on uncertainty estimation in the context of meteorites falling during dark flight.
Finally we publish the models for all the WRF runs we have calculated thus far.

\section{Input Data \& Methods} \label{sec:meth}

\subsection{Input Data} \label{sec:input}

Atmospheric conditions along the dark-flight trajectory are hindcast using the NCEP FNL (Final) Operational Model Global Tropospheric Analyses \citep[ds083.2;][]{NCEP2000}, archived by the National Center for Atmospheric Research (NCAR) Research Data Archive.
The FNL product provides global atmospheric analyses at $1^{\circ} \times 1^{\circ}$ horizontal resolution on different pressure levels from the surface to 10~hPa, at 6-hourly intervals.
Fields include horizontal wind components, temperature, relative humidity, and surface pressure.
These archived snapshots derived from the Global Data Assimilation System (GDAS) constrain the global atmospheric state at each time step and serve as boundary and initial conditions for the mesoscale model runs.

What observation data is part of these data products?
Surface observations, balloon data, wind profiler data, aircraft reports, buoy observations, radar observations, and satellite observations all contribute to the observations.
The FNL (final) data products are delayed so that more observational data can be used, with the idea that these hindcast data products should be more accurate than models used for forecast (forecasts, by definition, cannot use future observations or wait for late arriving data).

The FNL data products have one snapshot every 6 hours, at fixed times: 00z, 06z, 12z, 18z.
For WRF model initializatione download the four snapshots prior to the fall time and one after the fall time.

\subsection{WRF Preprocessing System}
The FNL data products cannot be used directly by the WRF solver.
We use the standard WRF Preprocessing System (WPS) of \citet{SkamarockEtAl2019}.
At a high level, these are the steps:
\begin{enumerate}    
    \item \texttt{geogrid}: defines the model grids on the map projection and interpolates static, time-invariant fields onto them.
    
    \item \texttt{ungrib}: reads the FNL GRIB2 files and writes decoded meteorological fields to WPS intermediate binary format.

    \item \texttt{metgrid}: horizontally interpolates the ungribbed meteorological fields onto the model grids defined by geogrid, and merges them with the static geographic fields.

    \item \texttt{real}: vertically interpolates from FNL's mandatory pressure levels to WRF's terrain-following sigma coordinate, and generates the initial and boundary condition files.
\end{enumerate}

\subsection{WRF Configuration and Integration}\label{sec:WRF}

Mesoscale atmospheric modelling is performed using the Weather Research and Forecasting (WRF) model version~4 with the Advanced Research WRF (ARW) dynamic core \citep{SkamarockEtAl2019}.
WRF is configured in hindcast mode: initialised from an archived FNL snapshot and propagated forward in time with forcing conditions constrained by subsequent archived snapshots bracketing the time of the meteorite fall.

Downscaling from the $1^{\circ}$ global grid to the resolution required for dark-flight modelling is achieved through a four-level nested domain configuration, with horizontal resolutions of 27~km, 9~km, 3~km, and 1~km (Tab. \ref{tab:wrf_domains}).
Each inner domain inherits its boundary conditions from the enclosing parent domain, with the innermost 1~km domain centered on the last known location of the meteorite (typically the last optically observed point  in bright flight).
This telescoping approach allows large-scale atmospheric dynamics from the global analysis to define realistic boundary conditions for the inner high-resolution domains.

It sometimes happens that the model crashes with the default configuration used (Tab. \ref{tab:wrf_domains}), usually in areas with complex geography (often near mountains and/or coastal areas).
When this happens we reduce the modelling time step to 1/4th of the default.
With earlier WRF versions we used smaller boundary area for the outer domain sizes, and not only reduced the time stamp, but also increased the domain size for problematic scenarios.

\begin{table*}[htbp]
  \centering
  \small
  \begin{tabular}{lcccccc}
    \toprule
    \textbf{Domain} &\textbf{ Horizontal cell size}  & \textbf{N cells} & \textbf{Extent (km)} & \textbf{Timestep (s)} & \textbf{Parent} & \textbf{Ratio} \\
    \midrule
    \textbf{d01} &  $27\,\mathrm{{km}}$ & $162 \times 162$ & $4374 \times 4374$ & 108 & d01 & $1:1$ \\
    \addlinespace[2pt]
    \textbf{d02} &  $9\,\mathrm{{km}}$ & $163 \times 163$ & $1467 \times 1467$ & 36 & d01 & $1:3$ \\
    \addlinespace[2pt]
    \textbf{d03} &  $3\,\mathrm{{km}}$ & $166 \times 166$ & $498 \times 498$ & 12 & d02 & $1:3$ \\
    \addlinespace[2pt]
    \textbf{d04} &  $1\,\mathrm{{km}}$ & $337 \times 337$ & $337 \times 337$ & 4 & d03 & $1:3$ \\
    \bottomrule
  \end{tabular}
  \caption{WRF default nested domain configuration}
  \label{tab:wrf_domains}
\end{table*}

\subsection{Model spin-up time}\label{sec:spinup}
Spin-up time is the period needed for the model's dynamical fields to become physically consistent after initialisation from coarser analysis data.
It is not clear from existing litterature what the ideal spin-up time should be for our use case (Req. \ref{req:model}), also considering that due to the meteorites descending relatively fast, we do not need to finely resolve shears and inversions.
The first paper that discusses spinup specifically for the WRF-ARW solver is the work of \citet{2007WtFor..22..501J}, who recommand a 12 h spinup time.
12 h and longer spin up times seem to have become standard practice \citep{2014BoLMe.152..213K}, however it is not clear from existing literature how detrimental shorter spinup times (e.g. 6 h) may be for our use case.
We have come across studies motivated by the observation of an extreme weather event (flooding typically) that was not predicted (or poorly predicted), effectively asking "what model parameters would have been able to predict this event?".
Some use multiple events as ground truth \citep{wes-7-1869-2022},
while some focus on a single event (e.g. \citet{LIU2023129443}) which may be prone to overfitting.
Anyhow the model parameters that produce the best fidelity of these extreme weather are of limited interest for our use case.

Hence we have run models ranging from almost no spinup time up to about 30 h, keeping in mind that short spinup times may be problematic. We then discuss the issue in a data-driven fashion, by looking at how close real meteorite fall cases are found compared to predictions (Sec. \ref{sec:val_met}).

\subsection{Computing Infrastructure and Performance}

We use the \textit{Setonix} supercomputer to do both the pre-processing steps, and run the WRF modelling.
Each model is run on one node, and each node is configured to use 64 or 128 CPUs.
With the default domain and time step, models run $\sim$4 times faster than real time.
This means that the longer run (initialised $\sim$ 24h prior to time of interest) takes about 6 hours to complete, in parallel of shorter models that become available earlier.
This turn around time is generally quick enough for the meteorite use case.
Unless a finer resolution is required to get a successful run,
running the models for one event burns through $\sim$ 1K core hours on the \textit{Setonix} supercomputer\footnote{\textit{Setonix} was commissioned in 2022 with AMD EPYC 7763 CPUs}.

\section{Results} \label{sec:results}

As part of this paper, we openly data release 1107 models from 302 unique events.
These are both from our archival data from the Global Fireball Observatory, and from more recent runs with the goal to have a near-complete record for post-2010 events, including: 
\begin{itemize}
    \item Meteorite falls observed by the Global Fireball Observatory (whether they were recovered or not);
    
    \item Weather Radar detection: \citet{2025AdAst202541760F} list in a number of meteorite falls and space debris re-entries observed by the US NEXRAD weather radar network, published online\footnote{\url{https://ares.jsc.nasa.gov/meteorite-falls/}, last accessed 2026-05-01};
    
    \item Instrumentally recorded and recovered meteorite falls from the 75 cases of \citet{2025M&PS...60..928J}, expanded with a couple more recent cases that have been reported online;
    
    \item Imminent impactors that landed on or near land \citep{2009Natur.458..485J,2021M&PS...56..844J,2024PSJ.....5..253K,2025NatAs...9.1624E,2024A&A...686A..67S,2026AcAau.240....1S}.
\end{itemize}

\begin{figure}
    \centering
    \includegraphics[width=\linewidth]{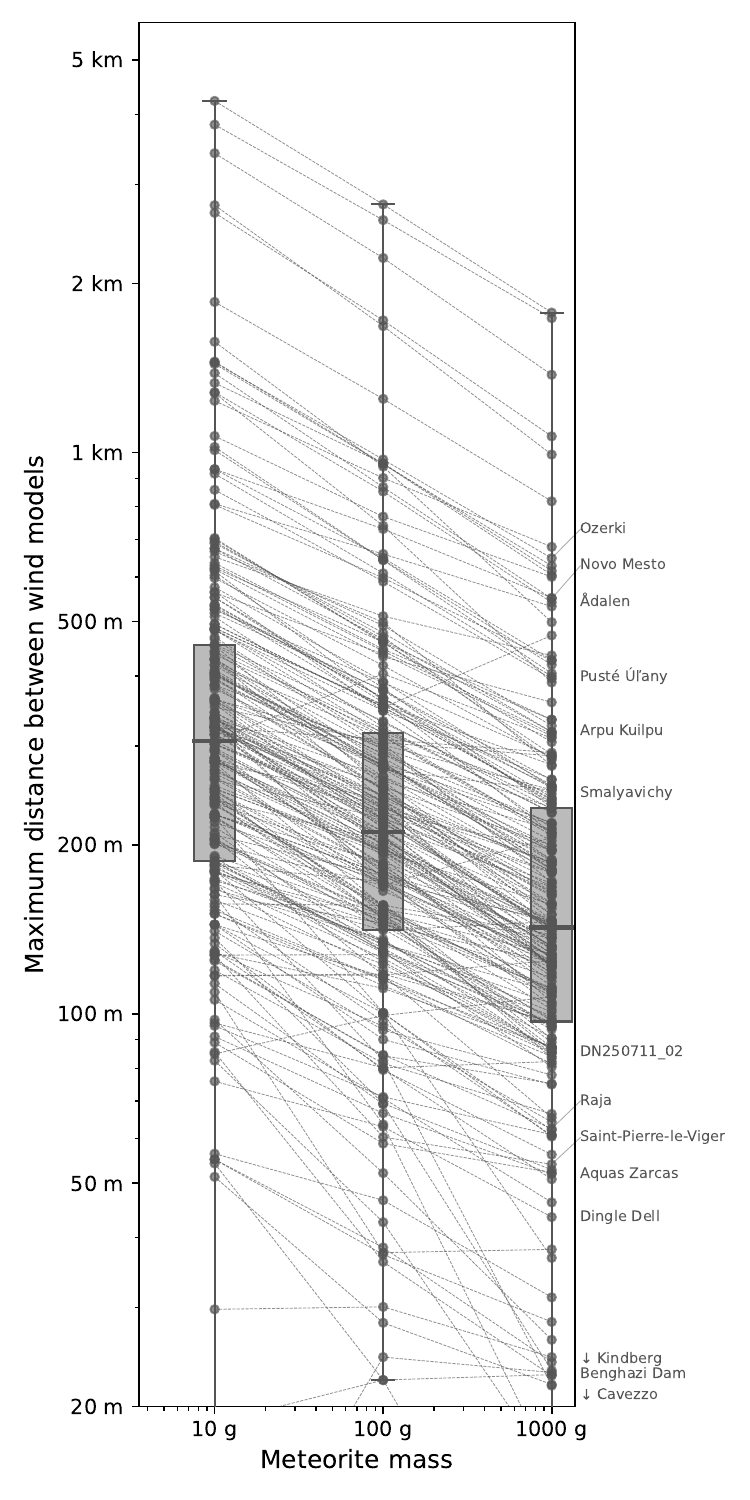}
    \caption{Maximum ground distance offset between hypothetical vertically dropped meteorites from 30 km altitude, using WRF models initialised at different times (excluding the WRF shortest run).
    Box plot represents Q1/median/Q3, with the extremities extending to mininimum/maximum.}
    \label{fig:max_distance}
\end{figure}

For each event, we attempt to start four WRF runs, corresponding to the four FNL snapshots prior to the event time.
If all runs fail for one event, we change the domain configuration following Sec. \ref{sec:WRF} until we get at least one of the models to run.

We then run a series of analyses on the atmospheric models we calculated, in order to understand their reliability and limitations.
To compare wind models calculated for a particular event to see how the difference would affect the dark flight of a meteorite, we run a series of dark flight simulations using the open source code of \citet{2022PSJ.....3...44T}.
To be able to compare different areas,
we used the same initial parameters for all events:
vertical fall from a height of 30 km at 1 km\,s$^{-1}$ speed.
For meteorite physical parameters, we adopt a density of 3,500 kg\,m$^{-3}$ (typical of ordinary chondrites), and run masses of 10, 100, and 1000 g.
For each event, we calculate the maximum ground distance between models for a given mass (Fig. \ref{fig:max_distance}).
As a comparison between events, we define for each event its "unreliability score" as the maximum ground difference between WRF models for a 1 kg meteorite, excluding models that are likely to suffer from spin-up issues (started $<$6\, before time of interest $t$).
Using the data from Fig. \ref{fig:max_distance}, we determine the relationship that links meteorite positional uncertainty as a function of mass.
Apart from a handful of outliers the general trend is a slope around -0.16.
Practically, this slope means that the unreliability score, defined above for a 1 kg meteorite, needs to be roughly doubled when looking at a 10 g meteorite.

\section{Discussion} \label{sec:disc}

Externally validating the accuracy of our models is difficult.
Publicly available ground truth data that would enable such validation would typically come from balloon radiosondes or wind profiling radars.
However these resources are part of the observations that make the input data files for the model (Sec. \ref{sec:input}).

In the sections below we attempt to validate some of our models using the locations of where past meteorites were found,
and also look at estimating uncertainty by internally comparing model runs started with different input data.

\subsection{Validation with where meteorites were found}\label{sec:val_met}

Here we detail successful meteorite recoveries that used the WRF modelling described in this paper:
\begin{itemize}
    \item Benghazi Dam \citep{2022LPICo2678.2888D}: 
Along with Cavezzo and Kindberg, the fall of Benghazi Dam is one of the top cases of meteorites recovered where WRF modelling is stable (unreliability score of 23 m).
Benghazi Dam meteorites were recovered thanks to rain radar returns propagated to the ground from 8.6 km altitude.
The first rock was found in under 10 minutes of searching,
indicating the accuracy of both radar-derived calculations, and the high fidelity of the wind models fed into the dark flight model.

\item Murrili \citep{2020M&PS...55.2157S}.
Murrili was used as an example case for the open-source dark flight integrator of \citet{2022PSJ.....3...44T}, who compare various WRF models in their Fig. 6.
 A clear outlier is a 3\,km resolution run started at 2015-11-26T18z (t-17h), which predicts the meteorite to be $>400$\,m from where it was found (northernmost line in Fig. 6 in \citet{2022PSJ.....3...44T}).
 Here we re-do their analysis, comparing all the WRF runs that we have calculated (Fig. \ref{fig:murrili_winds}), we find that its 1 km resolution counterpart is much closer to ground truth (Fig. \ref{fig:murrili_winds}).
All 1 km resolution models predict meteorites within $<200$\,m of where Murrili was found.
Apart from t-23h run, these solutions are somewhat unsatisfactory and are mostly inconsistent with each other, as evidenced by the unreliability score of 435 m.
This may be indicative of unstable weather that the models struggle to capture.
A major bushfire several hundred kilometres to the South started two days before the event may have contributed to this instability (large bushfires are known to create its own weather system \citep{2020NHESS..20.1497N}).

    \item Dingle Dell \citep{2018M&PS...53.2212D}:
The three wind models (started at t-24 h, t-6 h, and t-0 h) give near-identical results, and the meteorite was found 100 m from the resulting fall line.
Models started at t-18 h and t-12 h differ significantly and clearly not accurate \citep[Fig. 10 of][]{2018M&PS...53.2212D}.
It is difficult to explain why the mid-length model runs, that tend to give good results on other falls, fail to match the find.
It is also worth noting that the short 12z run, that only had minutes to spin-up, is a good match.

    \item Arpu Kuilpu \citep{2022M&PS...57.1146S}: The authors (we) did not discuss the results of different wind models in much detail.
A re-analysis shows that three of the models match the find well, as the meteorite was found 15\,m (t-10 h), 60\,m (t-2 h), and 110\,m (t-16 h) away from the fall line.
The longest run (t-22 h) gives a position 480m off from the ground truth.
Arpu Kuilpu came from the East, and made a complete U-turn in darkflight due to a $\sim20$\,m\,s$^{-1}$ jet stream; all four models disagree on the strength and/or the direction of this jet stream.
t-16, t-10, and t-2 h agree on the direction (West), but disagree on the strength (Fig. \ref{fig:arpu_profiles}): this creates a displacement along the East-West fall line, but operationally is does not significantly increase the search area.
Conversely, the t-22 h (12z) model predicts a South-West direction for the jet stream instead of West, causing the large shift to the North that we see on the ground.
If we track the evolution of this model over time, the t-22 h model starts diverging well before the event, and not just at jet stream altitude (Fig. \ref{fig:arpu_profiles}).
The 00z FNL checkpoint coincides with the synoptic time for most radiosonde launches in Australia.
This was likely the last time forcing observational data was input, as the 06z FNL checkpoint appears to have little influence over the model.
This is likely a typical example of model drift due to a too long runtime.

\begin{figure*}
    \centering
\includegraphics[width=1.\linewidth]{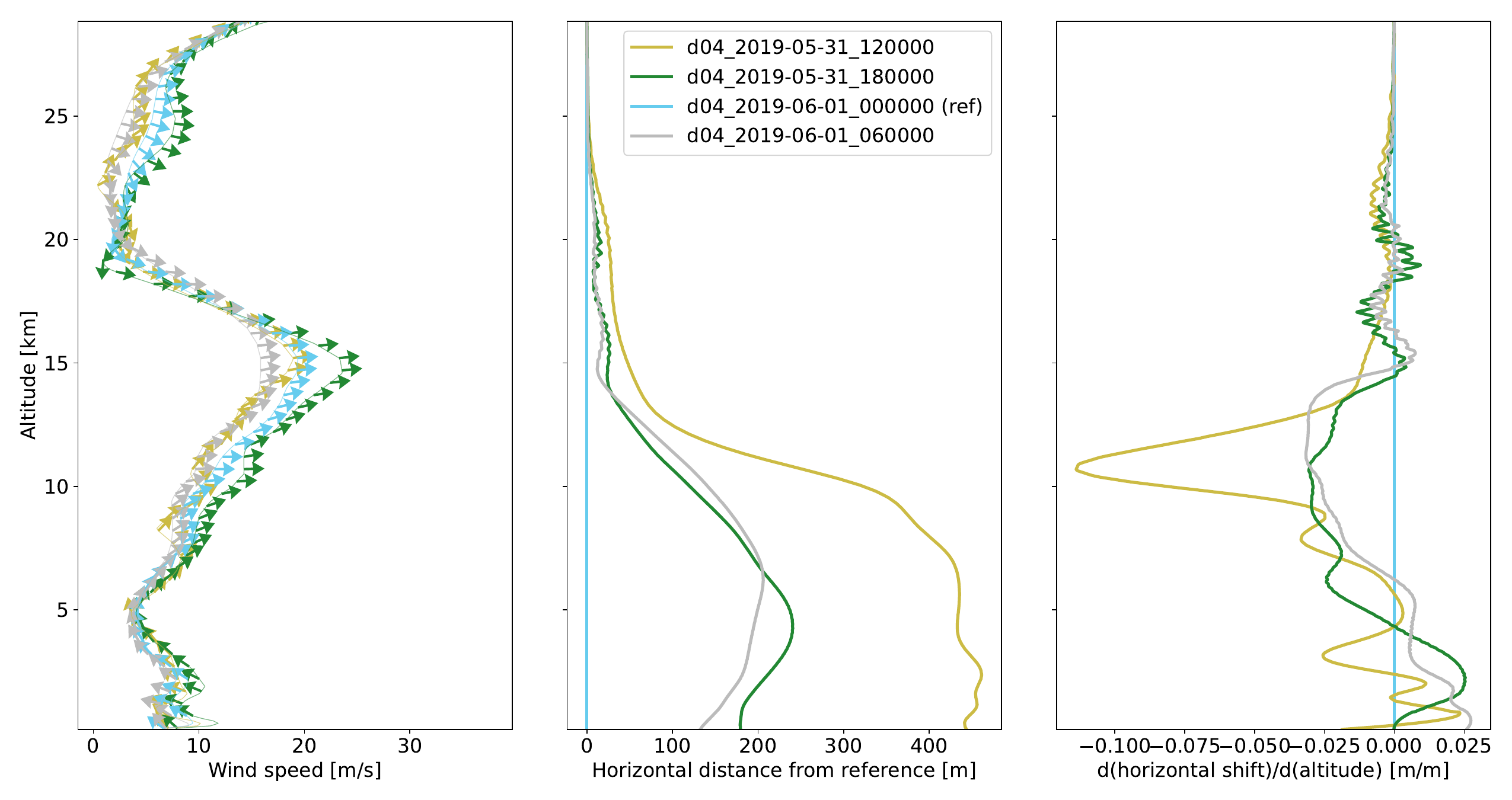}
    \caption{
    Vertical wind profiles comparison between WRF models for the Arpu Kuilpu meteorite fall.
    Left: wind speeds with arrows indicating direction.
    Centre: Models applied to the meteorite in dark flight (matching the found meteorite mass, density and ejection state vector): choosing the model that best fits the found meteorite as reference, we note the horizontal positional difference as the meteorite is falling, to understand where the major shifts happen.
    Right: gradient of the centre plot.
    While all four models are arguably all different, the feature that makes the most difference on the ground is the 12z predicting a South-West jet stream (9-13 km altitude), while the other three models predict a Westerly directly.
    The differences between 18z, 00z, and 06z on the ground are significant (100-200 m), but they are aligned along the fall line \citep[Fig. 8 of][]{2022M&PS...57.1146S}, a favourable case where area to be search does not increase.}
    \label{fig:arpu_profiles}
\end{figure*}

\begin{figure}
    \centering
\includegraphics[width=.9\linewidth]{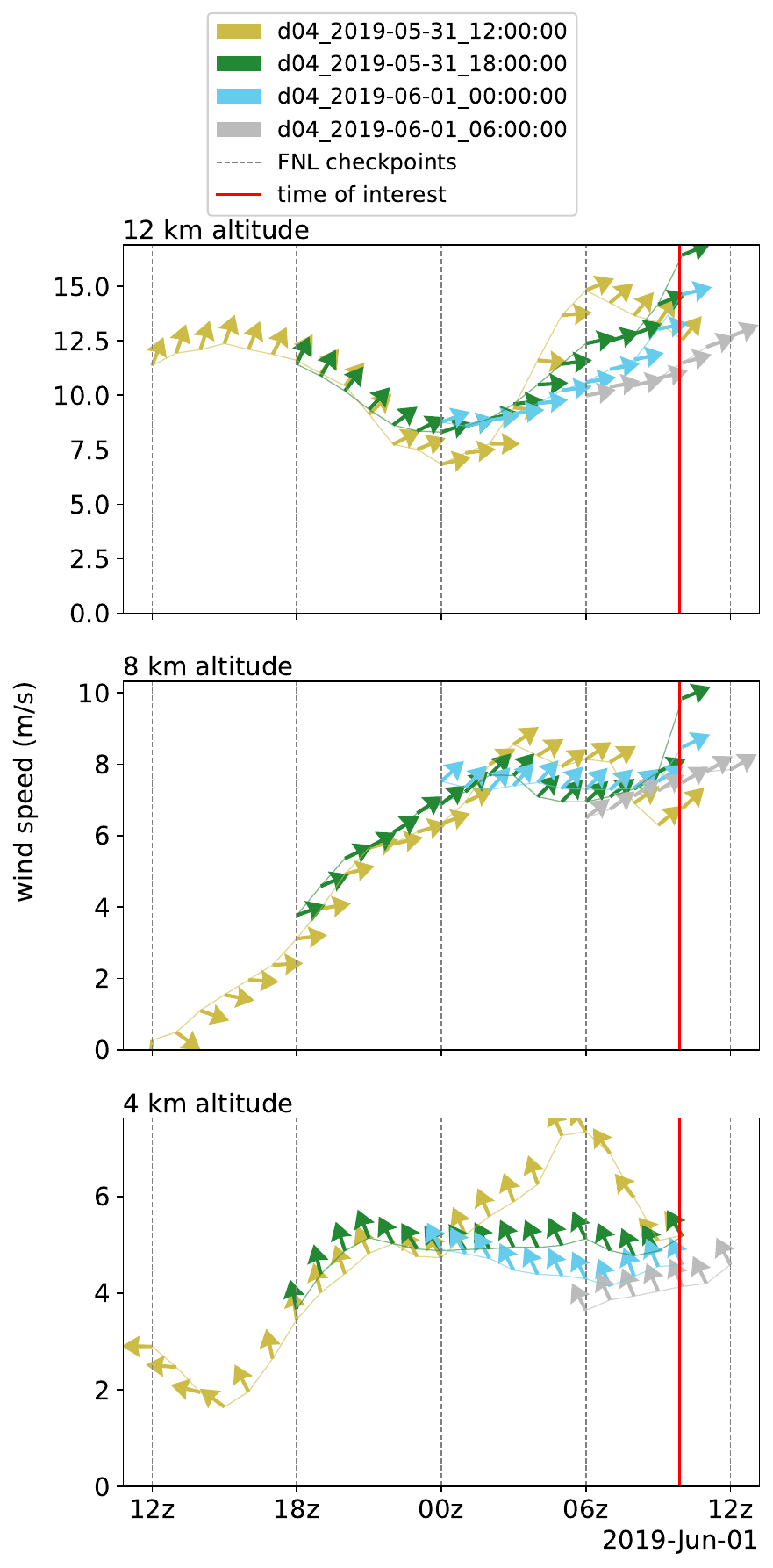}
    \caption{WRF models wind speed and direction tracking for the \textit{Arpu Kuilpu} meteorite.
    The 12z model (gold) is by far the worst match for the meteorite found on the ground. The mismatch mostly comes from the difference in wind speed in the jet stream layer, (seen here at 12 km altitude).
    }
    \label{fig:var_tracking}
\end{figure}

    \item Madura Cave \citep{2022M&PS...57.1328D}. The optical observations were of very poor quality (distant viewpoints, bad convergence angles).
The meteorite was found far from the nominal trajectory. During further analysis post-recovery, we realised that the nominal trajectory had to be off by over a kilometre, and we used the meteorite's location to infer what the likely valid ensemble of trajectories were.
Hence we cannot learn much about wind model accuracy with this case.

    \item Winchcombe \citep{2024M&PS...59..927M}. 
For Winchcombe the situation is complex.
Although the trajectory of the bolide is very well constrained overall (to about 60\,m), one viewpoint shows a vector change at the end, but the new vectors could not be calculated because this effect was only resolved by one camera.
Two large meteorites were recovered 300-400\,m North of the nominal fall line, significantly farther than what could be explained by the vector change at the end of bright flight (200 m).
The small fragments matched the dark flight simulations well.
At the time the analysis on Winchcombe was done only using the lower resolution (3 km) WRF runs that we had calculated at the time, and showed an acceptable internal consistency (reliability score 120\,m).
We have since rerun four new WRF models at high resolution (1 km), these paint a different picture:
Unreliability score is now 213 m.
The two longer runs give quite different results.
The t-22 h model is a perfect match ($<10$\,m) for the two large masses for Winchcombe, while t-16h is 150m off.
On the other hand, t-10h and t-4h cluster well with the results from the 3 km runs, giving results that appear correct for the small fragments, but that are significantly off for the larger masses.

    \item Kybo-Lintos -- first ever meteorite found with a drone \citep{2022ApJ...930L..25A,2026arXiv260519179A}. 
This fall is similar in observation circumstances as Madura Cave: distant viewpoints that do not form a good convergence angle.
Before the search we ran a Monte Carlo analysis on the possible ensemble of trajectories, which led to a search that encompasses a wide area of the fall line \citep[Fig. 1 of][]{2022ApJ...930L..25A}.
The meteorite was found very close to the nominal fall line though.
This may be suggesting that we overestimated astrometric uncertainties, but it could also very well be coincidental.
Hence we cannot learn much about this case.
The original dark flight simulations used 3 km resolution wind models, these are in good agreement with the more recent 1 km resolution runs, all consistent in this case.

    \item Taghzout \citep{2024LPICo3036.6030C}. 
All three WRF models that were run for this case (t-3h, t-9h, and t-15h) are very consistent: inter-model differences would not have "moved" the meteorite by more than 0.1 km.
However the meteorite was found $\sim$0.6 km from the nominal predicted fall line, but this was well within the 1$\sigma$ uncertainty area due to trajectory uncertainty (the trajectory solution relies on distant non dedicated viewpoints).
Hence we cannot learn much about wind field accuracy in this case.

    \item Al-Khadhaf \citep{2026M&PS...61..522Z}.
Uncertainty on the trajectory is difficult to estimate from the residuals as there were only two viewpoints.
The good quality astrometry points to a robust trajectory solution nonetheless.
Two small meteorites were found for Al-Khadhaf: 14 g and 8 g.
The 14\,g mass matches the fall line for both the 06z and 18z (t-14h and t-2h) WRF runs to within $<$50 m.
The 12z (t-8h) run, although farther from the find, is still a reasonable match ($\sim$ 110 m).
The 8\,g mass was found right on the edge of the 2$\sigma$ search area defined prior to the search \citep{2026M&PS...61..522Z}.
It does not match any of the fall lines, indicating that it was likely ejected from higher up on the trajectory.

    \item Raja \citep{Zappatini:2026:raja}.
Optical observations of the bright flight trajectory for Raja are good: residuals \citep[Fig. 2 of ][]{Zappatini:2026:raja} indicate that the  the high-resolution imagers produced a trajectory precise to $\sim$50\,m.
The four WRF models are very consistent (including the shortest run started at t-5h), and would have "moved" Raja by less than 100\,m.
Raja was found $<100$\,m from the predicted fall line.
We note that Raja was found a long way (0.75\,km) along the fall line, this cannot be explained by wind field uncertainty.

    \item Pindarri Punju Puri \citep{Clemente:2026:mdm}. The optical observations were of poor quality (distant viewpoints, bad convergence angles).
Meteorites of similar masses were found 1.7 km apart laterally, indicating an unmodelled vector change at the end of bright flight was likely the dominant effect at play.
Hence we cannot learn much about wind model accuracy with these cases.

    \item DN251107\_02\footnote{this meteorite is yet to be named}  \citep[fireball described by][]{2026PASA...43...27G} was found 310 m from the nominal fall line, despite an unreliability score of only 85 m.
Overall the astrometry and the resulting trajectory were of quality. However we could not astrometrically resolve individual fragments after the last major fragmentation point, which may have caused a late lateral impulse on the fragments, and may explain why the meteorite was found this far off the nominal fall lines.
Hence for this case we are reasonably confident that weather modelling  inaccuracy is not the likely explanation for the mismatch.
This will be discussed in detail in a forthcoming paper.

\end{itemize}

\begin{deluxetable*}{llllp{6cm}}
\tablecaption{Summary of which WRF models match the position of meteorites found.\label{tab:validation}}
\tablehead{\\
  \colhead{Meteorite} & \colhead{Unreliability score} & \colhead{Good fits} & \colhead{Poor fits} & \colhead{Note}
}
\startdata
Benghazi Dam        & 23\,m  & all      & ---                 &  \\
Murrili             & 435 m & t-29h, t-23h, t-17h, t-11h, t-5h & 3 km t-17h   & Complex case, see discussion.\\
Dingle Dell         & 44\,m  & t-24h, t-6h, t-0h & t-18h, t-12h & Unusual: mid-length runs are wrong, while both short and long runs are good. \\
Arpu Kuilpu         & 318\,m & t-10h, t-2h, t-16h & t-22h         & Jet stream direction mismatch in longest run. \\
Madura Cave         & 157\,m & ---   & ---             & Cannot assess: uncertain trajectory. \\
Winchcombe          & 213\,m & t-22h, t-16h & \multicolumn{1}{p{1.2in}}{t-16h, t-10h, t-4h,   all 3-km models} & Caveat: assessed for large masses only.  \\
Kybo-Lintos         & 107\,m & (all)  & ---             & Cannot assess: uncertain trajectory. \\
Taghzout            & 62\,m  & ---      & ---                 & Cannot assess: uncertain trajectory. \\
Al-Khadhaf          & 96\,m  & t-14h, t-2h & t-8h &  \\
Raja                & 63\,m  & all      & ---                 & \\
Pindarri Punju Puri & 136\,m & ---  & ---             & Cannot assess: uncertain trajectory. \& evidence of lateral kicks.  \\
DN251107\_02        & 85\,m  & ---  & (all)         & Likely due an unobserved lateral kick. \\
\enddata
\tablecomments{Unless noted otherwise, all models in this table are the high-resolution runs (1 km).}
\end{deluxetable*}

\subsection{How much does model variability affect meteorite fall locations?}\label{sec:variab}

Previous works have discussed using the different runs as a way to characterise uncertainty \citep{Osinski2020, Tewari2022} in WRF weather modelling.
For our WRF modelling for a given event, we start 3-4 model runs at the available FNL 6h data products in the t-30h to t-0h (Sec. \ref{sec:input}).
 We find that the situation varies widely between different events.
The median values (10\,g: 384\,m, 100\,g: 268\,m, 1\,kg:  195\,m) are poorly informative: the differences span over two orders of magnitude. 
If we exclude the short $<$6 h runs, we get differences that are appreciably smaller: 10\,g: 307\,m, 100\,g: 211\,m, 1\,kg:  143\,m (Fig. \ref{fig:max_distance}).
This indicates that in general spin-up is likely affecting shorter runs ($<6$\,h), even though the issue is not apparent in our limited validation set (Sec. \ref{sec:val_met}).

From an operational point of view, this exercise reveals that model differences need to be looked at on a case by case basis (Fig. \ref{fig:max_distance}).
Furthermore, our simplified dark flight runs do not take into account the specifics of each fall event.
The azimuth, slope, height, and speed of the fireball trajectory also play a major role in shaping a strewn field, and the combination of uncertainty on these variables work can work to increase or decrease the effective size of the search area (\textit{Arpu Kuilpu} is a typical example of this).

We discuss a selection of outlier events in which differences are very large.
The three "worst offenders" (top of Fig. \ref{fig:max_distance}) are in Australia, and all coincided with weather warning issued by the Bureau of Meteorology \footnote{past weather warning are not publicly accessible on the BOM website, but are findable on the Australian Broadcasting Corporation's weather section: \url{https://www.abc.net.au/news/topic/weather-warnings}, last accessed 2026-05-11}: \textit{DN160919\_02} ("severe weather warning for heavy rainfall in Western Victoria"), \textit{DN160623\_01} ("severe weather warning for damaging wind and heavy rainfall in South Australia"), and \textit{DN170113\_01} ("severe weather warning for heavy rainfall and damaging wind in South Australia").
They have unreliability scores of 1.8, 1.7, and 1.4 km, respectively.
We have not attempted to conduct meteorite searches for these events, but together they indicate that extreme and/or unstable weather is a major factor to WRF model unreliability.

This is confirmed by two other large outliers:
\begin{itemize}
\item Imminent Impactor 2024 RW1 on 2024-09-04z. %
Unreliability score 426 m.
In this case the weather unstability is likely explained by the nearby passage of Typhoon Yagi, that briefly reached Category 5 super typhoon status at 00:00 UTC on 5 September, just hours after the event \citep{2025MAP...137...21H}.

\item 2023-08-02 Kingsport NEXRAD event \footnote{\url{https://ares.jsc.nasa.gov/meteorite-falls/events/kingsport-tn-02}, last accessed 2026-05-05}. 
Unreliability score 395 m.
The reason why models diverge so much for this event is likely explained by extreme and unstable weather, with an Enhanced Fujita scale 1 class tornado in the same area the very next day \citep{TennesseeClimateOffice2023Aug}.\\
\end{itemize}

The top 10\% of outliers also include two documented fireballs for which meteorites have been recovered:

\paragraph{Murrili} Already discussed in Sec. \ref{sec:val_met}.

\begin{figure*}
    \centering
    \includegraphics[width=\linewidth]{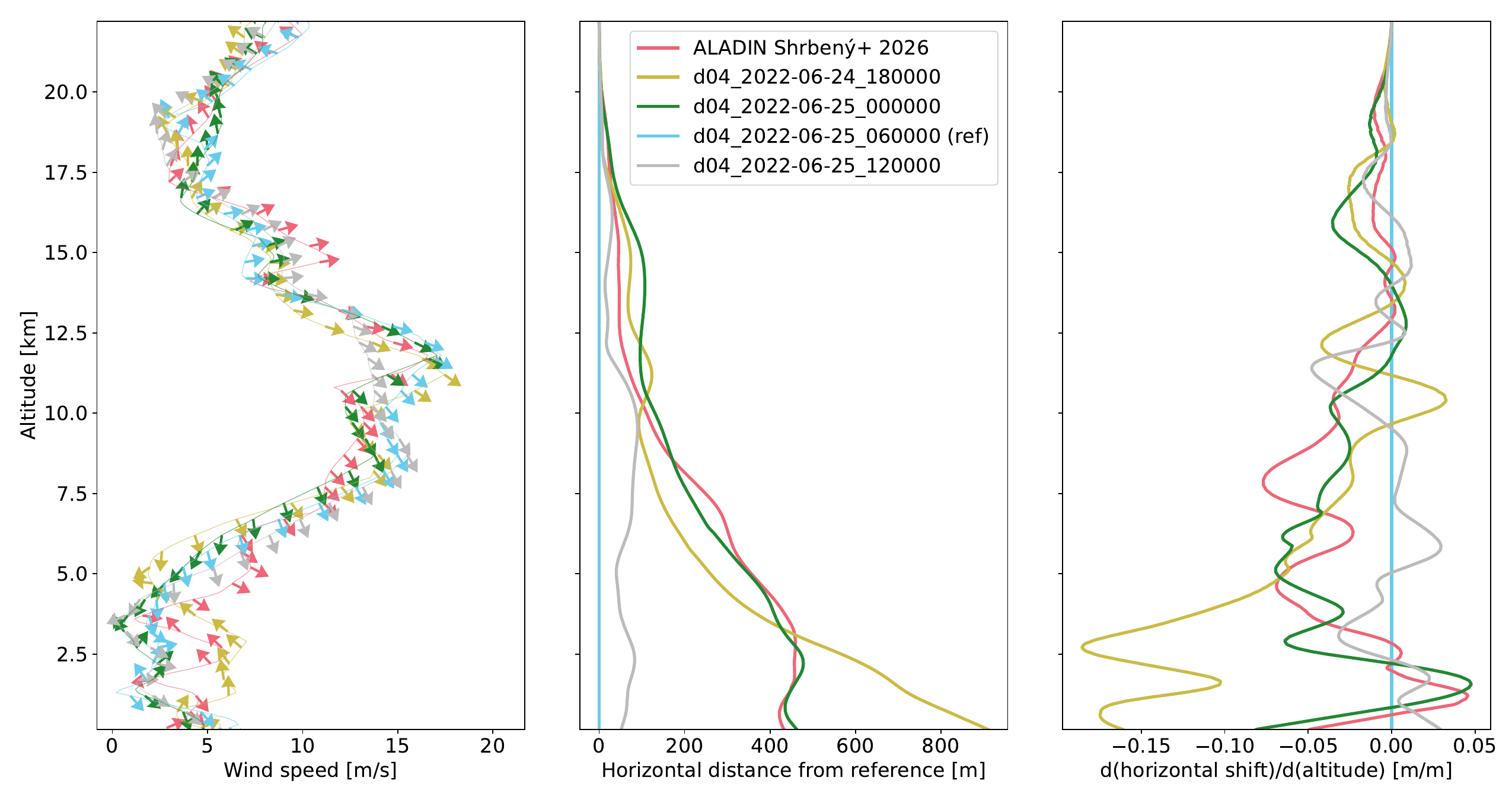}
    \caption{Vertical wind profiles comparison between WRF models and the Aire Limitée Adaptation Dynamique Développement International (ALADIN) weather model digitised from \citet{Shrbeny:2026}, for the Pusté Úľany meteorite fall.
    Left: wind speeds with arrows indicating direction.
    Centre: Models applied to the meteorite in dark flight (matching the found meteorite mass, density and ejection state vector): choosing the model that best fits the found meteorite as reference, we note the horizontal positional difference as the meteorite is falling, to understand where the major shifts happen.
    Right: gradient of the centre plot.
    The 00z and ALADIN models are very similar; they have the strongest impact on shifting the meteorite off course between 10 and 5 km altitude.
    The long run (started ar 18z the day before) diverges from other models from 10 km to the ground.
    Our reference (06z) and the 12z model give very close predictions; however such a close match on the ground (52 m) is partly coincidental, as the two virtual meteorites were 90 m apart at 9.6 km altitude: model differences throughout the column cancelled out in this case.
    }
    \label{fig:puste_models}
\end{figure*}

\begin{figure}
    \centering
    \includegraphics[width=\linewidth]{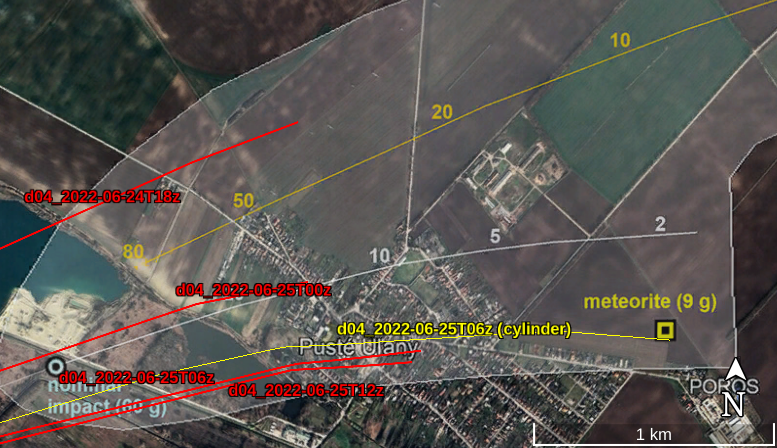}
    \caption{New fall lines for the Pusté Úľany meteorite fall.
    Red lines correspond assuming a spherical shape (eastern extremities correspond to $\sim$5 g), with four WRF models (labels correspond to initialisation time). Two of the models match the find quite well laterally. Along the fall line, a higher drag shape (cylinder option in the code of \citet{2022PSJ.....3...44T}) provides a somewhat better fit (yellow line). This figure is adapted from Fig. 8 of \citet{Shrbeny:2026}.}
    \label{fig:puste_fall_line}
\end{figure}

\paragraph{Pusté Úľany meteorite fall on 2022-06-25}
Unreliability score 399 m.
We have both meteorite ground truth \citep{puste_toth} and optical fireball trajectory.
\citet{Shrbeny:2026} discuss the circumstances of the fall in detail, and note two important things that are relevant here.
First, the meteorite was found quite far from the predictions published a posteriori: about 400 m South from nominal, $\gg$ than the $\sim$60 m residuals on the optical trajectory.
Second, the weather had just gone through a rapid change: "During the day, it was cloudy in most of our territory and it rained in many places, including local thunderstorms, but it started to clear up in the afternoon and evening".
We rerun dark flight simulation from the last fragmentation event with our four WRF wind models.
Two of the WRF models (started at t-6h and t-12h) fit the meteorite found well laterally (Fig. \ref{fig:puste_fall_line}).
The longer running models and the model used by \citet{Shrbeny:2026}, do not match well (Fig. \ref{fig:puste_models}).
Model drift issues with longer runs are well-known problems for limited-area hindcast simulations.
The chaotic nature of atmospheric flow means that small errors in the initial state grow exponentially with integration time, eventually saturating at scales that render interior solutions physically plausible but dynamically unrelated to the true atmosphere \citep{1969Tell...21..289L}.
Simultaneously, lateral boundary relaxation constrains only the outermost region of the domain, leaving mesoscale interior features free to diverge from reality in a way that worsens systematically with run length \citep{1997BAMS...78.2599W}.
Pusté Úľany is an example where model drift issues appear early, likely because of the sudden weather change described by \citet{Shrbeny:2026}.

\paragraph{Ådalen iron meteorite fall\citep{2026arXiv260215440M}}
Unreliability score 541 m.
Dark flight simulations by \citet{2026arXiv260215440M} predict the meteorite should have been 440 m from the actual impact point of where it was recovered.
Although this case is an outlier from a WRF modelling point of view, the a priori assumption is that a moderate inaccuracy in the wind model used would do little to offset a 13.8 kg iron lump.
To confirm this we run a darkflight, using our WRF models, and also predict the meteorite should be $\sim$0.5 km further North to where it was found.
The maximum distance between our four models is 270 m for Ådalen, not nearly enough to explain the discrepancy.
Hence the hypothesis of \citet{2026arXiv260215440M} of lift or steering forces at play is a better explanation for the mismatch.

\subsection{Effect of model spatial resolution on the results}\label{sec:res}
The majority of the model WRF runs we have calculated have been with 1 km resolution for the integration, and output at the same resolution.
Could we get away with using lower spatial resolution models, to save computing time?
While lower resolution models (Tab. \ref{tab:wrf_domains}) are calculated alongside the target 1 km resolution, the WRF integration feeds back the results into the lower resolution domains (WRF configuration flag \texttt{feedback = 1}), hence it makes no sense to compare two different resolutions from the same run.

\begin{figure}
    \centering
    \includegraphics[width=\linewidth]{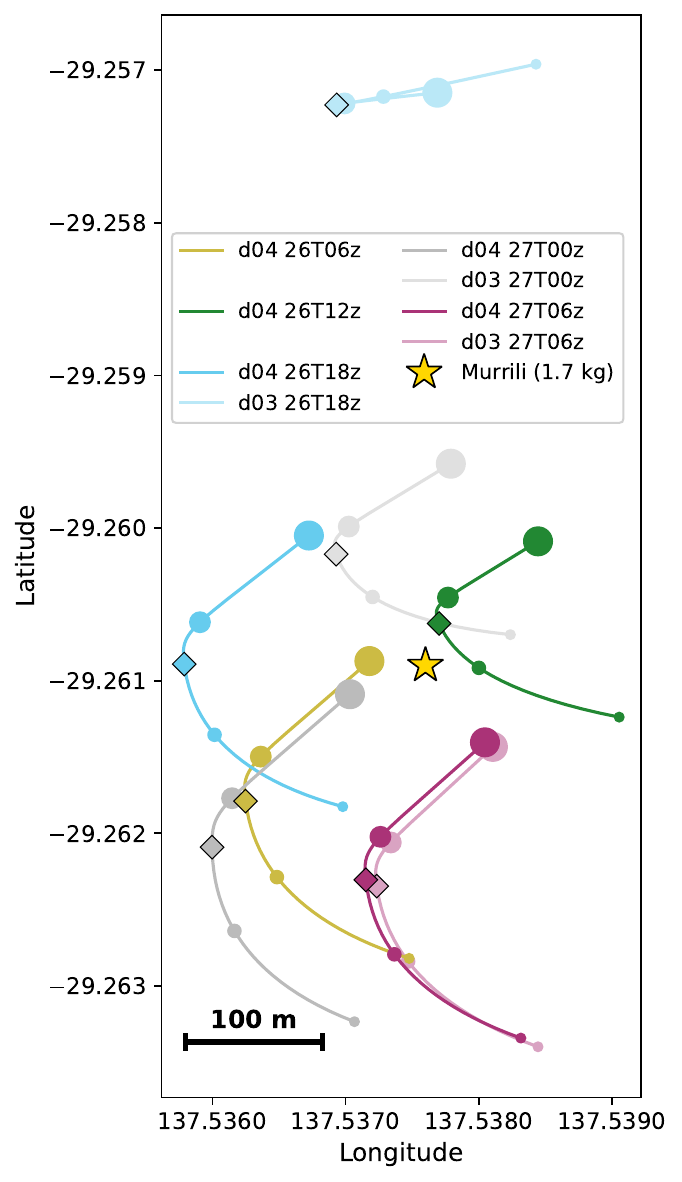}
    \caption{Effect of wind model choice on Murrili, including low-resolution (3 km) models. Large masses fall lines (0.5 to 5 kg) are represented, along with the fall location of Murrili. It is hard to objectively say which model fits the meteorite best.
    The "d03 26T18z" low-resolution model (light blue) is clearly off though, indicating that in this case the higher resolution version of this run (blue) is of higher fidelity.}
    \label{fig:murrili_winds}
\end{figure}

To compare wind models calculated with a  different spatial resolution, we have calculated both the 1 km (default) and 3 km resolution data products as two separate runs for a limited number of events.
For two runs started at the same time we have run synthetic darkflights of 10 g meteorites (same conditions as in Sec. \ref{sec:variab}), and noted the difference on the ground.
We get differences up to 499 m for \textit{2024 BX1 Ribbeck} (on the t-13h run), 309 m for Winchcombe (on the t-16h), 186 m for \textit{DN260327\_01} (on the t-13h run), 157 m for the \textit{Takapō} meteorite fall (t-14h run), 440 m for the \textit{Murrili} meteorite (on the t-17h run, see Fig. \ref{fig:murrili_winds}).
The selection of events for which we have both 1 and 3 km data products is limited, but enough to show that the differences are significant for our use case.

Which resolution is better?
Both Winchcombe (Tab. \ref{tab:validation}) and Murrili (Fig. \ref{fig:murrili_winds}) each show at least one WRF run where the high-resolution run matches the meteorites found significantly better than the low-resolution equivalent.
This seems to indicate that higher resolution runs yield better dark flight fidelity, although it is hard to draw a firm conclusion from just two ground truth cases.
The full analysis of the \textit{Ribbeck} and \textit{Takapō} meteorite falls in forthcoming papers should be able to confirm this trend.

\section{Conclusion}

\begin{enumerate}
\item Comparing WRF models wind variables provides an indication of solution robustness, and can be used to inform uncertainty bounds on the dark flight trajectory.
The median differences for a 1 kg meteorite is 143 m, doubling to 307 m for a 10 g meteorite.
Though we stress that model differences vary widely between events (two orders of magnitude on ground positions differences).%

\item Models started too shortly before the time of interest can suffer from spin-up issues.

\item Longer runs, although more stable physically and numerically, eventually suffer from model drift caused by early initialisation with very different forcing data.

\item It is hard to derive hard rules about which model start epoch to trust.
Users should make a critical assessment on a case by case basis, comparing dark flight results using the different wind models.

\item Our tests tend to indicate that using higher spatial resolution modelling (1 km instead of 3 km) improves the accuracy of dark flight predictions.

\item While it is clear that current weather observation datasets used with state-of-the-art hindcast tools are unable to meet our idealised requirement Req. \ref{req:model}, our approach is the first systematic approach to quantify dark flight uncertainty due to the wind field.

\item Looking towards the future, it is unlikely that a significant improvement in precision/accuracy happens unless the density of observation (radiosonde or wind profilers) increases. Triggering a radiosonde release at a nearby existing automated launch site when a meteorite fall is detected optically could be a relatively cheap way to increase dark flight fidelity.
\end{enumerate}

\newpage
We hope that these data products will be of use for:
\begin{enumerate}[label=\alph*)]

\item \textbf{Predicting the fall location of meteorites} from bright flight (fireball) observations \citep{2020P&SS..19105036D}, radar \citep{2025AdAst202541760F}, or even from pre-impact telescope observations \citep{2025Icar..42516345C}.

\item \textbf{Improving  meteoroid trajectory modelling}:  Meteoroid and meteorite trajectory modelling is an active area of research, with multiple sources of uncertainty and systematic issues.
With now over 80 meteorite falls instrumentally recorded and recovered \citep{2025M&PS...60..928J}, we are at a point where collectively this dataset can start informing where model systematic issues lie.
The present work is a step towards a better understanding of the reliability and the errors coming from the wind modelling, which has a significant impact on the dark flight. Future works may be able to focus on other sources of systematic errors.

\item \textbf{Bolide Acoustics}:
Bolide infrasound and seismic signals can be used to constrain source energy, trajectory, and fragmentation, but these inferences depend on acoustic ray tracing through a realistic atmosphere.
Winds refract and filter infrasound signals \citep{2026GeoRL..5320042S}, and fine-scale wind shear can produce multi-path arrivals that mimic fragmentation signatures \citep{2026Icar..45017007C}, making an accurate full-column wind specification a prerequisite for reliable bolide infrasound analysis \citep{2026GeoRL..5320188S}.

\item \textbf{Space Debris Re-entries}: Debris from re-entering spacecraft is a growing problem, becoming a danger to people and infrastructure on the ground, as well as a potential atmospheric pollution issue \citep{doi:10.1073/pnas.2313374120}.
Using similar principles as for meteorite falls, these models can be used to predict the strewn field from space debris re-entries \citep{Hatty:2026}.

\item \textbf{Sample Return Capsules}: Return capsules, both from interplanetary space (e.g. \textit{Hayabusa} missions) and from low-Earth orbit (e.g. \textit{Varda Space Industries W-Series}) have radio beacons and transponders that normally make locating them easy.
It is however conceivable that these primary location means may fail, in which case the ability to model their entry path from a previously known state vector would require a high-fidelity atmospheric model, especially for the under parachute phase of the flight.
\end{enumerate}

\newpage
\section{Data Release} \label{sec:data_release}

We publish the WRF models we have calculated thus far.
In order to limit the file size to permit the distribution of these files, we crop the models in time to only keep the previous time output before the event time, and all timesteps following the event.
We keep the entire calculated spatial domain.

The cropped netCDF data files, along with 1D vertical profile are published on the \textit{Zeonodo} platform: \citet{Devillepoix200_atmospheric,
Devillepoix2010_atmospheric,
Devillepoix2011_atmospheric,
Devillepoix2012_atmospheric,
Devillepoix2013_atmospheric,
Devillepoix2014_atmospheric,
Devillepoix2015_atmospheric,
Devillepoix2016_atmospheric,
Devillepoix2017_atmospheric,
Devillepoix2018_atmospheric,
Devillepoix2019_atmospheric,
Devillepoix2020_atmospheric,
Devillepoix2021_atmospheric,
Devillepoix2022_atmospheric,
Devillepoix2023_atmospheric,
Devillepoix2024_atmospheric,
Devillepoix2025_atmospheric,
Devillepoix2026_atmospheric,
Devillepoix_atmospheric_reentries}.
Our intention is to keep publishing new models in the future, and update the Zenodo records accordingly.
A simple way to discover what models are available is to use the frontend table/map we have developed \footnote{\url{https://desertfireballnetwork.github.io/freo_doctor}}.

\begin{acknowledgements}

This work was enabled by the Australian Research Council as part of the Australian Discovery Project scheme (DP170102529, DP200102073, DP230100301), and receives institutional support from Curtin University.
This work was supported by software support resources awarded under the Astronomy Data and Computing Services (ADACS) Merit Allocation Program. ADACS is funded from the Astronomy National Collaborative Research Infrastructure Strategy (NCRIS) allocation provided by the Australian Government and managed by Astronomy Australia Limited (AAL).
This work was supported by resources provided by the Pawsey Supercomputing Research Centre’s \textit{Setonix} Supercomputer, with funding from the Australian Government and the Government of Western Australia.

A Large Language Model was used for: creating code to archive and upload the dataset to Zenodo, assist with creating plots for the paper, as well as for developing the front-end webpage. The text in the paper is entirely human original writing.

\texttt{Freo Doctor}: The Fremantle "Freo" Doctor is the name of the afternoon sea breeze in Boorloo/Perth (Western Australia), a major contributor to the author's wind-powered hobbies.

\end{acknowledgements}

\bibliography{main}
\bibliographystyle{aasjournalv7}

\end{document}